\documentclass[11pt]{article}
\usepackage{amsmath,amssymb,bbm,amsthm}
\usepackage{fullpage}
\usepackage{thm-restate,color,xcolor,xspace}
\usepackage{hyperref,cleveref,graphicx}
\usepackage{algorithm,algorithmicx}
\usepackage[noend]{algpseudocode}

\newcommand{\f}{\frac}
\newcommand{\cd}{\cdot}

\newcommand{\sr}{\sqrt}

\newcommand{\lds}{\ldots}

\newcommand{\sm}{\setminus}
\newcommand{\s}{\subseteq}
\newcommand{\su}{\supseteq}

\newcommand{\BE}{\begin{enumerate}}
\newcommand{\EE}{\end{enumerate}}
\newcommand{\im}{\item}
\newcommand{\BI}{\begin{itemize}}
\newcommand{\EI}{\end{itemize}}
\def\BAL#1\EAL{\begin{align*}#1\end{align*}}
\def\BALN#1\EALN{\begin{align}#1\end{align}}
\def\BG#1\EG{\begin{gather}#1\end{gather}}

\newcommand{\logn}{\log n}

\newcommand{\inv}{^{-1}}

\newcommand{\R}{\mathbb R}

\newcommand{\e}{\epsilon}
\newcommand{\de}{\delta}

\newcommand{\la}{\lambda}

\newcommand{\pt}{\partial}
\newcommand{\al}{\alpha}

\newcommand{\Om}{\Omega}
\newcommand{\el}{\ell}

\newcommand{\m}{\mathcal}

\newcommand{\lf}{\lfloor}
\newcommand{\rf}{\rfloor}
\newcommand{\lc}{\lceil}
\newcommand{\rc}{\rceil}

\newcommand{\polylog}{\textup{polylog}}
\newcommand{\pl}{\textup{polylog}}

\newcommand{\lp}{\left(}
\newcommand{\rp}{\right)}

\newcommand{\lmt}{\left[\begin{matrix}}
\newcommand{\rmt}{\end{matrix}\right]}

\newtheorem{theorem}{Theorem}[section]
\newtheorem{lemma}[theorem]{Lemma}
\newtheorem{definition}[theorem]{Definition}
\newtheorem{corollary}[theorem]{Corollary}
\newtheorem{observation}[theorem]{Observation}
\newtheorem{claim}[theorem]{Claim}
\newtheorem{subclaim}{Subclaim}
\newtheorem{fact}[theorem]{Fact}
\newtheorem{assumption}[theorem]{Assumption}

\newcommand{\BT}{\begin{theorem}}
\newcommand{\ET}{\end{theorem}}
\newcommand{\BL}{\begin{lemma}}
\newcommand{\EL}{\end{lemma}}
\newcommand{\BD}{\begin{definition}}
\newcommand{\ED}{\end{definition}}
\newcommand{\BC}{\begin{corollary}}
\newcommand{\EC}{\end{corollary}}
\newcommand{\BO}{\begin{observation}}
\newcommand{\EO}{\end{observation}}
\newcommand{\BCL}{\begin{claim}}
\newcommand{\ECL}{\end{claim}}
\newcommand{\BSCL}{\begin{subclaim}}
\newcommand{\ESCL}{\end{subclaim}}
\newcommand{\BF}{\begin{fact}}
\newcommand{\EF}{\end{fact}}
\newcommand{\BA}{\begin{assumption}}
\newcommand{\EA}{\end{assumption}}
\newcommand{\BP}{\begin{proof}}
\newcommand{\EP}{\end{proof}}
\newcommand{\BSP}{\begin{subproof}}
\newcommand{\ESP}{\end{subproof}}
\newcommand{\BPS}{\begin{proof}[Proof (Sketch)]}
\newcommand{\EPS}{\end{proof}}
\Crefname{observation}{Observation}{Observations}
\Crefname{claim}{Claim}{Claims}
\Crefname{subclaim}{Subclaim}{Subclaims}
\Crefname{fact}{Fact}{Facts}
\Crefname{assumption}{Assumption}{Assumptions}

\newenvironment{subproof}[1][\proofname]{%
  \begin{proof}[#1]%
}{%
  \end{proof}%
}

\newcommand{\alert}{\textcolor{red}}
\newcommand{\para}{\paragraph}

\newcommand{\tO}{\tilde{O}}

\newcommand{\thml}[1]{\label{thm:#1}}
\newcommand{\thm}[1]{\Cref{thm:#1}}
\newcommand{\leml}[1]{\label{lem:#1}}
\newcommand{\lem}[1]{\Cref{lem:#1}}
\newcommand{\defnl}[1]{\label{def:#1}}
\newcommand{\defn}[1]{\Cref{def:#1}}
\newcommand{\clml}[1]{\label{clm:#1}}
\newcommand{\clm}[1]{\Cref{clm:#1}}
\newcommand{\corl}[1]{\label{cor:#1}}
\newcommand{\cor}[1]{\Cref{cor:#1}}

\newcommand{\eqnl}[1]{\label{eq:#1}}
\newcommand{\eqn}[1]{(\ref{eq:#1})}
\newcommand{\linel}[1]{\label{line:#1}}
\renewcommand{\line}[1]{line~\ref{line:#1}}
\newcommand{\secl}[1]{\label{sec:#1}}
\renewcommand{\sec}[1]{\Cref{sec:#1}}

\newcommand{\bd}{\mathbf{d}}

\begin{document}

\title{Deterministic Minimum Cut in Poly-logarithmic Maximum Flows\thanks{A preliminary version of this paper appeared in the Proceedings of the IEEE Annual Symposium on Foundations of Computer Science (FOCS), 2020.}}
\author{
Jason Li\thanks{This work was done as a graduate student at Carnegie Mellon University.}\\Simons Institute for Theory of Computing \\ UC Berkeley\\Email: {\tt jmli@cs.cmu.edu}
\and 
Debmalya Panigrahi\\Department of Computer Science\\Duke University\\Email: {\tt debmalya@cs.duke.edu}
}
\date{}

\maketitle
\begin{abstract}
    We give a deterministic algorithm for finding the minimum (weight) cut of an undirected graph on $n$ vertices and $m$ edges using $\polylog(n)$ calls to any maximum flow subroutine. Using the current best deterministic maximum flow algorithms, this yields an overall running time of $\tilde O(m \cdot \min(\sqrt{m}, n^{2/3}))$ for weighted graphs, 
    and $m^{4/3+o(1)}$ for unweighted (multi)-graphs. This marks the first improvement for this problem since a running time bound of $\tO(mn)$ was established by several papers in the early 1990s.
    
    Our global minimum cut algorithm is obtained as a corollary of a minimum Steiner cut algorithm, where a minimum Steiner cut is a minimum (weight) set of edges whose removal disconnects at least one pair of vertices among a designated set of terminal vertices. The running time of our deterministic minimum Steiner cut algorithm matches that of the global minimum cut algorithm stated above. Using randomization, the running time improves to $m^{1+o(1)}$ because of a faster maximum flow subroutine; this improves the best known randomized algorithm for the minimum Steiner cut problem as well.

    Our main technical contribution is a new tool that we call {\em isolating cuts}. Given a set of vertices $R$, this entails finding cuts of minimum weight that separate (or isolate) each individual vertex $v\in R$ from the rest of the vertices $R\setminus \{v\}$. Na\"ively, this can be done using $|R|$ maximum flow calls, but we show that just $O(\log |R|)$ suffice for finding isolating cuts for any set of vertices $R$. We call this the {\em isolating cut lemma}.

\end{abstract}



\section{Introduction}
\label{sec:introduction}
The minimum cut (or min-cut) of an undirected, weighted graph
$G = (V, E, w)$ is a minimum weight subset of edges
whose removal disconnects the graph.
Finding the min-cut of 
a graph is one of the central problems in combinatorial 
optimization, dating back to the work of 
Gomory and Hu~\cite{gomory1961multi} in 1961 who gave
an algorithm to compute the min-cut of an $n$-vertex
graph using $n-1$ max-flow computations. 
Since then, a large body of 
research has been devoted to obtaining faster algorithms 
for this problem. In 1992, Hao and Orlin~\cite{hao1992faster}
gave a clever amortization of the $n-1$ max-flow computations
to match the running time of a single max-flow 
computation. Using the ``push-relabel'' max-flow algorithm 
of Goldberg and Tarjan~\cite{GoldbergT88}, they obtained 
an overall running time of $O(mn \log (n^2/m))$ on 
an $n$-vertex, $m$-edge graph. 
However, their amortization technique is specific to the 
push-label algorithm, and cannot be applied to
faster max-flow algorithms that have been designed since their work
(e.g., by Goldberg and Rao~\cite{goldberg1998beyond}). 
Around the same time, Nagamochi and Ibaraki~\cite{nagamochi1992computing}
(see also \cite{NagamochiI92})
designed an algorithm that bypasses max-flow computations 
altogether, a technique that was further refined by
Stoer and Wagner~\cite{stoer1997simple} (and independently
by Frank in unpublished work).
This alternative method yields a running time of 
$O(mn + n^2 \log n)$. Prior to our work, these works yielding a
running time bound of $\tO(mn)$ were the fastest 
{\em deterministic} min-cut algorithms for weighted graphs.

Starting with Karger's contraction algorithm in 1993~\cite{karger1993global}, 
a parallel body of work started to emerge in {\em randomized} algorithms
for the min-cut problem. This line of work (see also Karger and Stein~\cite{karger1996new})
eventually culminated in a breakthrough paper by Karger~\cite{Karger00}
in 1996 that gave an $O(m\log^3 n)$ time {\em Monte Carlo} algorithm 
for the min-cut problem. Note that this algorithm comes to within 
poly-logarithmic factors of the optimal $O(m)$ running time for 
this problem. In this paper, Karger asks whether we can also achieve
near-linear running time using a 
{\em deterministic} algorithm. Even before Karger's work, Gabow~\cite{gabow1995matroid} 
showed that the min-cut can be computed in $O(m + \lambda^2 n\log (n^2/m))$
(deterministic) time, where $\lambda$ is the value of the min-cut
(assuming integer weights). Note that this result obtains a 
near-linear running time if $\lambda$ is a constant, but 
in general, the running time can be exponential. 

In a recent 
breakthrough, Kawarabayashi and Thorup~\cite{KT18}
gave the first near-linear time deterministic algorithm for
the min-cut problem in {\em simple graphs}. They obtained a running 
time of $O(m \log^{12} n)$, which was later improved 
by Henzinger, Rao, and Wang~\cite{HenzingerRW17} to 
$O(m \log^2 n\log\log^2 n)$. From a technical perspective, 
their work introduced the idea of using low conductance 
cuts to find the min-cut of the graph, a very powerful idea that 
we also exploit in this paper. Nevertheless, in spite of this exciting 
progress, the question of designing a faster deterministic 
min-cut algorithm for general weighted graphs (or unweighted
multi-graphs) remained open.

In this paper, we give the following result:
\begin{restatable}{theorem}{main}\thml{main}
Fix any constant $\e>0$.
There is a deterministic min-cut algorithm for weighted\footnote{For simplicity, all weights are assumed to be polynomially bounded throughout the paper.} undirected graphs that makes $\pl(n)$ calls to $s$--$t$ max-flow on a weighted undirected graph with $O(n)$ vertices and $O(m)$ edges, and runs in $O(m^{1+\e})$ time outside these max-flow calls.\footnote{The exponent in the polylog($n$) term denoting the number of max-flow computations is a function of $\e$.} If the original graph $G$ is unweighted, then the inputs to the max-flow calls are also unweighted. Using the current fastest deterministic max-flow algorithms on unweighted (multi-)graphs (Liu and Sidford~\cite{liu2020faster}) and weighted graphs (Goldberg and Rao~\cite{goldberg1998beyond}) respectively, this implies a deterministic min-cut algorithm for unweighted (multi-)graphs in $m^{4/3+o(1)}$ time and for weighted graphs in $\tilde O(m \cdot \min(\sqrt{m}, n^{2/3}))$ time.
\end{restatable}
This represents the first improvement in the running time 
of deterministic (or even Las Vegas) algorithms for the min-cut problem on
general (weighted/multi) graphs since the early 1990s. 
An advantage of our result is that unlike the algorithm of Hao and Orlin that 
relied on amortizing runs of a specific max-flow algorithm, our algorithm is agnostic to the specific max-flow algorithm being used. Therefore, our result will automatically improve as 
progressively better max-flow algorithms are discovered.

A classic generalization of the min-cut problem is the {\em Steiner}
min-cut problem. In this problem, we are given an undirected, 
weighted graph $G = (V, E, w)$ and a subset $T\s V$ of terminals.
The Steiner min-cut is a minimum weight subset of edges whose 
removal disconnects at least one pair of terminals in the graph. 
Note that this interpolates between the min-cut problem defined
above ($T = V$) and the $s-t$ min-cut problem ($T = \{s, t\}$).
For this problem on general, weighted graphs, the best known 
algorithm previous to our work was to perform $|T|-1$ max-flow 
computations by fixing a source vertex $s\in T$ and iterating
over all sink vertices $t\in T\setminus \{s\}$, and report 
the minimum weight cut among the $s-t$ min-cuts returned by 
these max-flow calls~\cite{DinitzV94}. For unweighted graphs, better algorithms
are known~\cite{ColeH03,HariharanKP07,BhalgatHKP08} with the 
best running time being $\tilde O(m + \lambda^2 n)$, where 
$\lambda$ is the value of the Steiner min-cut~\cite{BhalgatHKP07}.
In this paper, we give a randomized algorithm for the Steiner
min-cut problem in weighted graphs:
\begin{restatable}{theorem}{randsteiner}\thml{randsteiner}
There is a randomized Steiner min-cut algorithm for weighted undirected graphs that makes $O(\log^3n)$ calls to $s$--$t$ max-flow on a weighted undirected graph with $O(n)$ vertices and $O(m)$ edges, and runs in $O(m\log^2n)$ time outside these max-flow calls. If the original graph $G$ is unweighted, then the inputs to the max-flow calls are also unweighted. Using the current fastest (randomized) max-flow algorithms, this implies a (randomized) Steiner min-cut algorithm for weighted graphs that runs in $m^{1+o(1)}$ time using the recent $m^{1+o(1)}$-time max-flow algorithm of Chen~{\em et al.}~\cite{ChenKLPPS22}.
\end{restatable}

We also derandomize the Steiner min-cut algorithm to obtain a deterministic algorithm for the problem. In fact, our deterministic min-cut algorithm (\Cref{thm:main}) is obtained as a corollary of our deterministic Steiner min-cut algorithm given by the next theorem:
\begin{restatable}{theorem}{detsteiner}\thml{detsteiner}
Fix any constant $\e>0$.
There is a deterministic Steiner min-cut algorithm for weighted undirected graphs that makes $\pl(n)$ calls to $s$--$t$ max-flow on a weighted undirected graph with $O(n)$ vertices and $O(m)$ edges, and runs in $O(m^{1+\e})$ time outside these max-flow calls.\footnote{The exponent in the polylog($n$) term denoting the number of max-flow computations is a function of $\e$.} If the original graph $G$ is unweighted, then the inputs to the max-flow calls are also unweighted. Using the current fastest deterministic max-flow algorithms on unweighted (multi-)graphs (Liu and Sidford~\cite{liu2020faster}) and weighted graphs (Goldberg and Rao~\cite{goldberg1998beyond}) respectively, this implies a deterministic Steiner min-cut algorithm for unweighted (multi-)graphs in $m^{4/3+o(1)}$ time and for weighted graphs in $\tilde O(m \cdot \min(\sqrt{m}, n^{2/3}))$ time.
\end{restatable}

To obtain the above theorems, we introduce our main tool that we call {\em minimum isolating cuts}:
\BD[Minimum isolating cuts]
Consider a weighted, undirected graph $G=(V,E)$ and a subset $R\s V$ of size at least $2$. The \emph{minimum isolating cuts} for $R$ is a collection of sets $\{S_v:v\in R\}$ satisfying $S_v=\arg\min\{w(\pt S)\mid S\cap R=\{v\}\}$ for all $v\in R$. In other words, $S_v$ is (the side containing $v$ of) the minimum cut separating $v$ from $R\sm \{v\}$.  
\ED

Na\"ively, minimum isolating cuts can be computed by using $|R|$ max-flows by setting each vertex $v$ as the source vertex $s$ and connecting all the other vertices $R\setminus \{v\}$ to a common sink vertex $t$ with edges of infinite capacity. In this paper, we improve this na\"ive bound and give an algorithm for finding all minimum isolating cuts that uses $O(\log |R|)$ max-flow calls. We state this next:

\begin{restatable}[The Isolating Cut Lemma]{theorem}{isolating}\thml{isolating}
Fix a subset $R\s V$ of size at least $2$. There is an algorithm that computes the minimum isolating cuts for $R$ using $\lc\lg|R|\rc+1$ calls to $s$--$t$ max-flow on weighted graphs having $O(n)$ vertices and $O(m)$ edges, and takes $O(m\logn)$ deterministic time outside of the max-flow calls. If the original graph $G$ is unweighted, then the inputs to the max-flow calls are also unweighted.
\end{restatable}

\paragraph{Impact and Subsequent Results.} There has been a significant amount of research activity related to this paper since its first publication. In terms of results, Li~\cite{Li21} gave an $m^{1+o(1)}$-time deterministic algorithm for the min-cut problem in undirected graphs, thereby resolving the deterministic complexity of the global minimum cut problem. This algorithm uses a different set of techniques from this paper, and can viewed as a de-randomization of Karger's randomized near-linear time min-cut algorithm. In terms of techniques, the main tool introduced in this paper -- minimum isolating cuts -- has been shown to be useful for a broad variety of graph connectivity problems, some of which we outline below: 
\begin{itemize}
    \item For any $\epsilon > 0$, Li and Panigrahi~\cite{LiP21} used isolating cuts to obtain an algorithm for computing a $(1+\epsilon)$-approximate Gomory-Hu tree (and therefore, $(1+\epsilon)$-approximations of $s$--$t$ min-cut values for all vertex pairs $s, t$) of a weighted, undirected graph using poly-logarithmic max-flow calls. Prior to this work, no algorithm was known for the Gomory-Hu tree problem -- exact or approximate -- on general, weighted graphs that uses fewer than $n-1$ calls to a max-flow subroutine. 
    \item Li~{\em et al.}~\cite{LiNPSY21} adapted minimum isolating cuts to vertex cuts and obtained a vertex min-cut algorithm in unweighted graphs using poly-logarithmic max-flow calls. This was the first improvement for the problem in 25 years since the work of Henzinger, Rao, and Gabow~\cite{HenzingerRG96}.
    \item Abboud {\em et al.}~\cite{AbboudKT21a} independently developed a minimum isolating cuts subroutine, and used it in an exact algorithm for Gomory-Hu tree in simple graphs that runs in $\tO(n^{5/2})$ time. This running time was subsequently improved to $\tO(n^2)$ independently by Abboud {\em et al.}~\cite{AbboudKT21b}, Li~{\em et al.}~\cite{LiPS21}, and Zhang~\cite{Zhang21}. Finally, in \cite{AbboudKLPST21}, this result was generalized to arbitrary weighted graphs (from simple unweighted graphs) thereby achieving the first improvement for the Gomory-Hu tree problem in general, weighted graphs in 60 years since the work of Gomory and Hu in 1961. 
    \item Cen, Li, and Panigrahi improved the running time of edge connectivity augmentation and splitting off problems using the isolating cuts framework~\cite{CenLP22a}. This was then further improved to near-linear time in \cite{CenLP22b} using other techniques. 
    \item The isolating cuts framework also extends beyond graph cuts and applies to any symmetric, submodular function. This was observed by Chekuri and Quanrud~\cite{ChekuriQ21} and independently by Mukhopadhyay and Nanongkai~\cite{MukhopadhyayN21}, which leads to faster algorithms for finding minimizers in the context of vertex connectivity, element connectivity, and hypergraph cuts.
\end{itemize}

\subsection{Our Techniques} \secl{techniques}

At a high level, our algorithm combines two key insights originating from prior work on graph cut algorithms:
 \BE
 \im If the min-cut is an \emph{unbalanced} cut, then it should be susceptible to \emph{local} graph cut algorithms pioneered by Spielman and Teng~\cite{SpielmanT03,AndersenCL07,OrecchiaZ14,HenzingerRW17,NanongkaiSY19,nanongkai2019computing}. While our isolating cut lemma is not local (in the precise definition of \emph{local} in this line of work), it is nevertheless inspired by local graph algorithms.
 \im If the min-cut is a \emph{balanced} cut, then it must have low \emph{conductance}, which was exploited by the deterministic min-cut algorithm of Kawarabayashi and Thorup for \emph{simple} graphs~\cite{KT18}.
\EE

To incorporate both ideas simultaneously, our algorithm divides the problem into two cases: when some target min-cut is \emph{unbalanced}---for example, when one side of the cut has at most $\polylog(n)$ vertices---and when it is \emph{balanced}.

\para{Unbalanced Case and Minimum Isolating Cuts.}

Suppose we have a subset $R\s T$ of \emph{red} terminals, where $|R|\ge2$, with the following property: {\em one of the two sides of the min-cut intersects $R$ in {\bf exactly one} vertex}. In this ideal scenario, we can simply compute the minimum isolating cuts for $R$ and return the isolating cut of smallest weight, which is indeed the global min-cut. 

We now briefly describe our minimum isolating cuts algorithm that uses $O(\log|R|)$ max-flow computations. This algorithm has two phases. First, the algorithm computes $O(\log |R|)$ different bi-partitions of the $R$ vertices, such that each pair of vertices $u,v\in R$ is separated in at least one of the bi-partitions. Then, for each bi-partition $(A,B)$ of $R$, the algorithm computes a min-cut separating $A$ and $B$ using a max-flow subroutine. Now, imagine removing all the edges in (the union of) these $O(\log|R|)$ many min-cuts. This would split the graph into connected components, each containing at most one vertex in $R$; for each $v\in R$, define $C_v\s V$ as the vertices of this connected component. Let $S$ be the side of some min-cut containing a single vertex $v\in R$, and assume without loss of generality that $S$ is inclusion-wise minimal. Our key observation is that, by the submodularity of cuts, we must have $C_v\supseteq S$. In particular, if we contract $V\sm C_v$ in the original graph $G$ into a single vertex $t$, then $(S,(C_v\cup\{t\})\sm S)$ is still a min-cut of the same value. Therefore, it suffices to compute a $v$--$t$ min-cut in this contracted graph (through a single $v$--$t$ max-flow computation). Of course, we do not know which component $C_u$ contains the set $S$, so we try them all. But the fact that the components $C_u$ are disjoint means that the total number of vertices and edges over all these max-flow instances is $O(n)$ and $O(m)$, respectively. Therefore, the cumulative cost of these max-flow computations is equal to just a single max-flow computation in the entire graph. This concludes the minimum isolating cuts algorithm and the case when a side of the min-cut contains exactly one vertex in $R$.

Now, what happens if the set $R$ contains not a single vertex of a side $S$ of the min-cut, but $\polylog(n)$ vertices? If randomization were allowed, sub-sampling each vertex in $R$ with probability $1/\polylog(n)$ is sufficient. Fortunately, this random sampling can be de-randomized: there exists a deterministic construction of a family of $\polylog(n)$ subsets of $R$ such that some set $T\s R$ in this family is guaranteed to satisfy $|T\cap S|=1$ and $|T|\ge2$. 
The de-randomization procedure is standard and builds off the concept of \emph{splitters}~\cite{alon1995color}. We then apply the aforementioned algorithm on each subset of $R$ in this family (instead of on $R$ itself).

Finally, note that by setting $R=T$, we obtain an algorithm that correctly computes a Steiner min-cut through $\polylog(n)$ calls to max-flow, given that a min-cut exists with $\polylog(n)$ vertices on one side. This suffices for the unbalanced case.

\para{Balanced Case.}
Note that if randomization were allowed, the aforementioned random sampling procedure also handles the balanced case. In particular, if the smaller side $S$ of the target min-cut has around $n/2^i$ vertices for some $1\le i\le\lg n$, then if $R$ is a random sample of $2^i$ vertices, we still have $|S\cap R|=1$ with constant probability. By sampling $O(\logn)$ many subsets $R$ at each of the $\lf\lg n\rf$ scales $2^i$ (i.e., for each of the $\lf\lg n\rf$ integral values of $i$), our algorithm succeeds w.h.p.

The issue with de-randomization, however, is that very small sampling probabilities are difficulties to de-randomize. For example, if $|S|=\sr n$, then we are effectively sampling each vertex with probability $1/\sr{n}$ into $R$, which is much smaller than $1/\polylog(n)$. In this case, the deterministic construction would produce $(\sr n)^{O(1)}$ many subsets of $R$, which is too many.

For the balanced case, assume that each side $S$ of every min-cut satisfies $|S\cap R|>\polylog(n)$. Here, our solution is not to solve the min-cut outright, but to make ``progress'' in a different way: we ``sparsify'' $R$ by replacing it with a subset $R'\s R$ of at most half the size, such that if $R$ intersects both sides of some target min-cut in more than $\polylog(n)$ vertices, then $R'$ intersects both sides of the same min-cut in at least $1$ vertex. We begin the algorithm with $R=T$ and this sparsification of $R$ can be performed at most $O(\log n)$ times before $|R|\le \polylog(n)$.\footnote{The pseudocode in our main Algorithm~\ref{alg:main} actually names this set $U$ instead of $R$ to distinguish it from the set $R$ used as input to the algorithm at the beginning of the unbalanced case.} If, for each intermediate $R$, we always have $|S\cap R|>\polylog(n)$ for each side $S$ of some min-cut, then by the sparsification guarantee, the final $R$ still intersects both sides of some target min-cut. But since $|R|\le \polylog(n)$ now, we can simply iterate over all pairs $s,t\in R$ and compute a min $s$--$t$ cut in $G$ for each such pair. Of course, it might also happen that in an intermediate step, $|S\cap R| < \polylog(n)$, and we are in the unbalanced case at that stage. So, we also run the algorithm for the unbalanced case described earlier in each step of the sparsification procedure.

It remains to describe the sparsification procedure of $R$. For simplicity, we will only consider the case $R=T=V$ here, since the ideas remain the same for general $R$ and $T$ but the notation is more cumbersome. Here, our crucial observation, motivated by~\cite{KT18}, is that the smaller side $S$ of some target min-cut must have \emph{sparsity} $w(\pt S)/|S|$ at most $\la/\polylog(n)$, where $\la$ is the min-cut value. Fix a parameter $\phi=1/\polylog(n)$, and assume that $|S|\ge1/\phi^3$, which gives sparsity $w(\pt S)/|S|\le \phi^3\la$. We now compute an \emph{expander decomposition} of the graph, informally defined as follows: we partition the vertex set $V$ into $V_1,\lds,V_\el$ such that (1) each induced graph (or ``cluster'') $G[V_i]$ has no cuts of sparsity at most $\phi\la$, for a slightly extended notion of sparsity that we omit in this sketch, and (2) the total weight $w(E(V_1,\lds,V_\el))$ of edges across different clusters is at most (roughly) $\phi \la n$ (ignoring lower-order factors). Since each set $V_i$ induces a cut $(V_i,V\sm V_i)$ of weight at least $\la$, a simple counting argument shows that the number of clusters $\el$ is at most (roughly) $\phi n$.

Recall that the set $S$ has sparsity at most $\phi^3\la$, which is much smaller than the sparsity of any cut inside any cluster $G[V_i]$ (which must be at least $\phi\la$). Intuitively, this means that $S$ cannot cut too ``deeply'' into any cluster. In the ideal case, where $S$ does not cut into \emph{any} cluster (i.e., either $S\cap V_i=\emptyset$ or $S\s V_i$ for each $i\in[\el]$), then selecting an arbitrary vertex from each $V_i$ into $R'$ suffices: $R'$ intersects both sides $S$ and $V\sm S$ of the min-cut and satisfies $|R'|\lesssim\phi n\le n/2$. In general, $S$ may cut a little into each cluster, but we show that by adding a small, arbitrary selection of vertices from each $V_i$ into $R'$, we can still guarantee $|R'|\le n/2$ and ensure that $R'$ intersects both sides of the min-cut.

\section{Minimum Isolating Cuts}
\label{sec:single-intersect}
As mentioned in the introduction, one of our main algorithmic components is computing the \emph{minimum isolating cuts} for a subset $R$ of vertices, which is the focus of this section. As mentioned in the \textbf{Unbalanced Case} of \sec{techniques}, this immediately implies a Steiner min-cut algorithm given the additional input $R\s T$ with $|R|\ge2$, and under the promise that there exists a side $S$ of some Steiner min-cut satisfying $|S\cap R|=1$. We then handle the more general case $|S\cap R|\le\pl(n)$ in \sec{hash}.

We first introduce a few standard graph-theoretic definitions. For a graph $G=(V,E,w)$ and a subset $U\s V$ of vertices, define $\pt_GU$ as the set of edges of $G$ with exactly one endpoint in $U$; when the graph $G$ is clear from context, we drop the subscript $G$ and use $\pt U$ instead. For a subset $F\s E$ of edges, define $w(F):=\sum_{e\in F}w(e)$ as the total weight of edges in $F$. In particular, $w(\pt U)$ is the total weight of edges with exactly one endpoint in $U$.

Let us now formally define the minimum isolating cuts and the corresponding isolating cut lemma.
\BD[Minimum isolating cuts]
Consider a weighted, undirected graph $G=(V,E)$ and a subset $R\s V$ of size at least $2$. The \emph{minimum isolating cuts} for $R$ is a collection of sets $\{S_v:v\in R\}$ satisfying $S_v=\arg\min\{w(\pt S)\mid S\cap R=\{v\}\}$ for all $v\in R$. In other words, $S_v$ is (the side containing $v$ of) the minimum cut separating $v$ from $R\sm v$. 
\ED

\BT[Isolating cut lemma]\thml{local}
Fix a subset $R\s V$ of size at least $2$. There is an algorithm that computes the minimum isolating cuts for $R$ using $\lc\lg|R|\rc+1$ calls to $s$--$t$ max-flow on weighted graphs having $O(n)$ vertices and $O(m)$ edges, and takes $O(m\logn)$ deterministic time outside of the max-flow calls. If the original graph $G$ is unweighted, then the inputs to the max-flow calls are also unweighted.
\ET

\begin{figure}\centering
\includegraphics{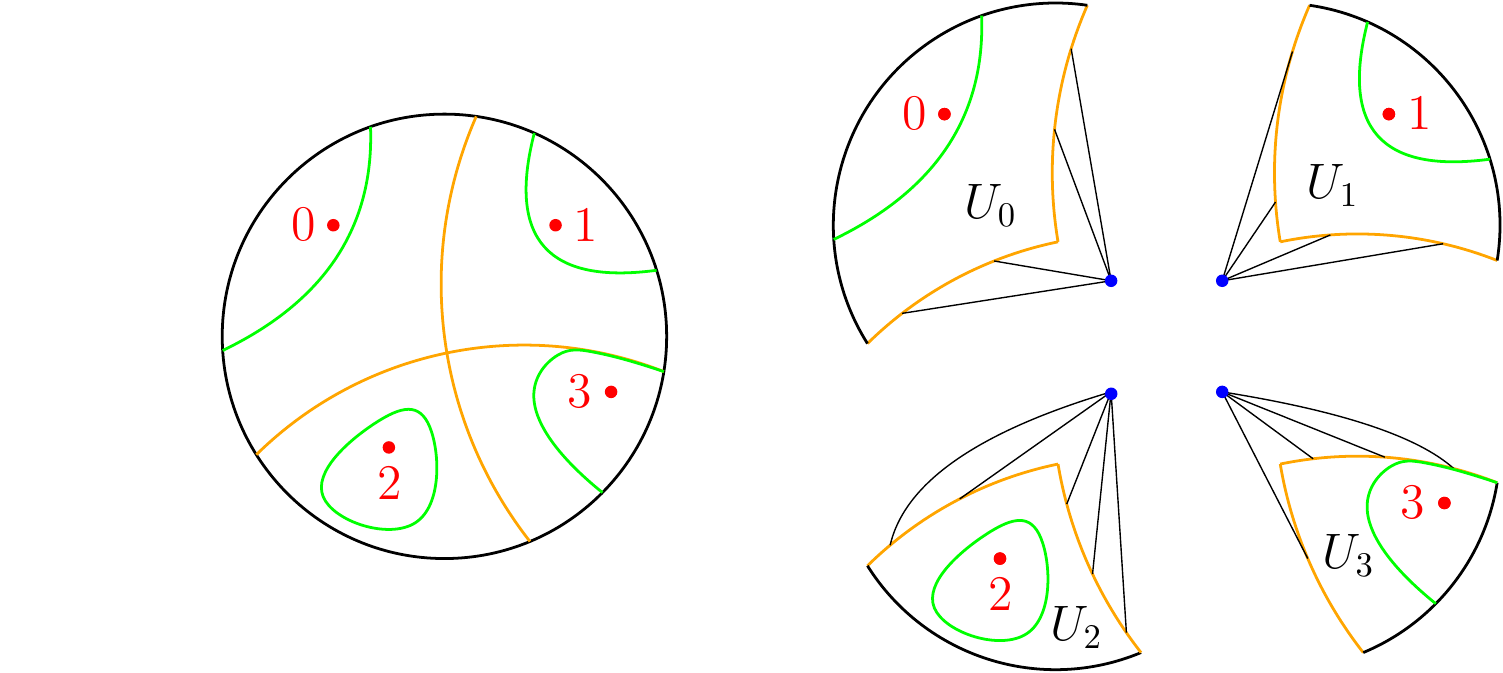}
\caption{The minimum isolating cuts algorithm for $|R|=4$. The orange marks the ``upper boundary'' of each green isolating cut. They are formed by the min-cut separating $\{0,1\}$ and $\{2,3\}$ and the min-cut separating $\{0,2\}$ and $\{1,3\}$.}\label{fig:1}
\end{figure}
Our main idea is to first compute, for each vertex in $R$, an ``upper boundary'' to the location of the min-cut separating that vertex from the rest of $R$ (see Figure~\ref{fig:1}). More precisely, for each vertex $v\in R$, we want to compute a set $U_v$ of vertices that contains $S_v$. If we can do so, then it suffices to compute an $(v,t)$-min-cut on the graph $G$ with $V\sm U_v$ contracted to a single vertex $t$, which will return $\pt S_v$ or some other $(v,R\sm v)$-min-cut. To make this min-cut computation fast, we would like $U_v$ to be small, ideally not much larger than $S_v$. We are not able to prove such a strong guarantee, but we can ensure that the sets $U_v$ are \emph{disjoint} among all $v\in R$, which suffices for our running time bound. 

Our procedure to compute the sets $U_v$ is as follows. We first compute $\lc\lg|R|\rc$ many bipartitions of $R$ such that any two vertices in $R$ are separated in at least one bipartition. For each bipartition $(A,B)$ of $R$, we compute the min-cut separating $A$ and $B$, and then for each vertex $v\in R$, we set $U_v$ as the common intersection of the sides containing $v$ of the $\lc\lg|R|\rc$ many computed min-cuts. We show by a simple submodularity argument that the side containing $v$ of each of the $\lc\lg|R|\rc$ min-cuts must contain $S_v$ (if we assume $S_v$ to be minimal in a sense), and thus, their common intersection $U_v$ also contains $S_v$.

The rest of this section formalizes the above intuition to prove \thm{local}.

\BP[Proof (\thm{local}).]
Order the vertices in $R$ arbitrarily from $1$ to $|R|$, and let the {\em label} of each $v\in R$ be its position in the ordering, a number from $1$ to $|R|$ that is denoted by a unique binary string of length $\lc\lg|R|\rc$.
Let us repeat the following procedure for each $i=1,2,\lds,\lc\lg|R|\rc$. For each vertex $v$, color it \emph{red} if the $i$'th bit of its label is $0$, and \emph{blue} if the $i$'th bit of its label is $1$. Then, compute a min-cut $C_i\s E$ in $G$ between the red vertices and the blue vertices (for iteration $i$).

For each vertex $v\in R$, let $\la_v$ be the minimum value of $w(\pt S)$ over all $S\s V$ satisfying $S\cap R=\{v\}$, and let $S^*_v$ be an inclusion-wise minimal set satisfying $S^*_v\cap R=\{v\}$ and $w(\pt S^*_v)=\la_v$. Also, define $U_v$ as the connected component in $G \sm \bigcup_i C_i$ containing $v$. 
\BCL\clml{v}
$|U_v\cap R|=\{v\}$ for all $v\in R$.
\ECL
\BP
By definition, $v\in U_v\cap R$. Suppose for contradiction that $U_v\cap R$ contains another vertex $u\ne v$. Then, there is a path $P$ in $U_v$ from $u$ to $v$. Since the binary strings assigned to $u$ and $v$ are distinct, they differ in their $j$'th bit for some $j$. Then, the cut $C_j$ must separate $u$ and $v$, and in particular, $P$ must contain at least one edge in $C_j$. But $P$ is contained in $U_v$, which is contained in $G \sm \bigcup_i C_i$, so $P$ cannot contain an edge in $C_j$, a contradiction.
\EP
\BCL\clml{Uv}
$U_v \supseteq S^*_v$ for all $v\in R$.
\ECL
\BP
Fix a vertex $v\in V$ and an iteration $i$. Let the side of the cut $C_i$ containing $v$ be $T_v^i\s V$; we claim that $S^*_v\s T_v^i$. Suppose for contradiction that $S^*_v\cap T_v^i \subsetneq S^*_v$; then by submodularity,
\[  w(\pt( S^*_v\cup T_v^i)) +  w(\pt(S^*_v\cap T_v^i)) \le  w(\pt S^*_v) +  w(\pt T_v^i) .\]
Since $S^*_v\cap T_v^i$ satisfies $S^*_v\cap T_v^i=\{v\}$ and $S^*_v\cap T_v^i\subsetneq S^*_v$, we must have $w(\pt(S^*_v\cap T_v^i))>\la_v= w(\pt S^*_v)$ by our choice of $S^*_v$ to be inclusion-wise minimal. Therefore,
\[  w(\pt( S^*_v\cup T_v^i)) <  w(\pt T_v^i)  .\]
But $(S^*_v\cup T_v^i)\cap R = T_v^i\cap R$, and in particular, the cut $\pt(S^*_v\cup T_v^i)$ also separates red vertices from blue vertices. This contradicts the choice of $\pt T_v^i=C_i$ as the min-cut separating red vertices from blue vertices.

Therefore, over all iterations $i$, none of the edges in the induced subgraph $G[S^*_v]$ are present in $C_i$. Note that $G[S^*_v]$ is a connected subgraph; therefore, it is a subgraph of the connected component $U_v$ of $G\sm\bigcup_iC_i$ containing $v$.
\EP

It remains to compute the desired set $S_v$ given the set $U_v\su S_v$. Starting from $G$, contract $R\sm U_v$ into a single vertex $t$; we want to compute the min $v$--$t$ cut in the contracted graph $G_v$, which corresponds to a set $S_v$ satisfying $S_v\cap R=\{v\}$ by \clm{v}. Since $\pt_{G_v}S^*_v$ is a valid $v$--$t$ cut in this graph by \clm{Uv}, we have $w(\pt_{G_v} S_v)\le w(\pt_{G_v}S^*_v)=w(\pt _GS^*_v)=\la_v$, as desired.

Note that each edge in $E$ is either in exactly one graph $G_v$, or it is adjacent to $t$ in exactly two graphs $G_v$. Therefore, the total number of edges over all graphs $G_v$ is at most $2m$. We can compute $v$--$t$ min-cut on all $G_v$ in ``parallel'' through a single max-flow call on the disjoint union of all $G_v$. Note that if the original graph $G$ is unweighted, then this max-flow instance is also unweighted. Finally, recovering the sets $S_v$ and the values $w(\pt S_v)$ take time linear in the number of edges of $G_v$, which is $O(m)$ time over all $v\in R$.

This completes the proof of \thm{local}.
\EP

\section{Randomized Algorithm for Minimum Steiner Cut}
\label{sec:rand-mincut}

In this brief section, we note that the minimum isolating cuts algorithm of \thm{local} easily implies a randomized Steiner min-cut algorithm that makes $\pl(n)$ many calls to max-flow. 

\randsteiner*
The algorithm essentially calls \thm{local} $O(\log^2 n)$ times; on each iteration, $R\s T$ is a random set of vertices sampled at a particular scale.

For each positive integer $i\le\lg n$, repeat the following procedure $O(\logn)$ times: let $R\s T$ be a random sample of $2^i$ vertices, and call \thm{local} on the set $R$ to obtain a cut $S_v$ for each $v\in R$. Return the cut $S_v$ with the minimum value of $w(\pt S_v)$ over all $v$ and over all the iterations.

We claim that w.h.p., the returned cut $S_v$ is a global min-cut. Let $S^*$ be the smaller side of the global min-cut. Observe that if, in any iteration, the sampled set $R$ satisfies $|R\cap S^*|=1$, then \thm{local} will find the global min-cut. Consider the integer $i=\lf \lg (n/|S^*|) \rf$. Then, for each iteration where $R$ is a random sample of size $2^i$, we sample exactly one vertex in $S^*$  with probability $\Om(1)$. Since we sample at this scale $O(\logn)$ times, this occurs at least once w.h.p.

\section{Deterministic Algorithm for Minimum Steiner Cut}
\label{sec:det-mincut}
In this section, we present our deterministic min-cut algorithm and prove our main result, \thm{detsteiner}, which is restated below:

\detsteiner*

\begin{figure}\centering
\includegraphics{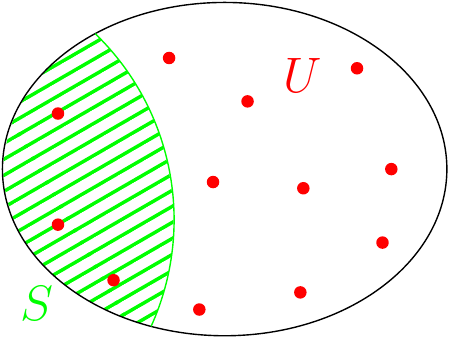}
\caption{$U$ is $k$-unbalanced for $k=3$.}\label{fig:2}
\end{figure}

Our high-level idea (see \Cref{alg:main}) is essentially to de-randomize the random selection of vertices in $R$. Our main tools will be constructions of hash families and expander decomposition. Throughout the algorithm, we maintain a set $U\s T$ of vertices that starts out as $U=T$ and shrinks over time. We distinguish between the cases when $U$ is \emph{$k$-unbalanced} or \emph{$k$-balanced} for some $k=\pl(n)$, as defined below~(see Figure~\ref{fig:2}).

\BD[$k$-unbalanced, $k$-balanced]\defnl{balanced}
For any positive integer $k$, a subset $U\s V$ is \emph{$k$-unbalanced} if there exists a side $S\s V$ of some min-cut satisfying $|S\cap U|\le k$. More specifically, we say that $U$ is \emph{$k$-unbalanced with witness $S$}. The subset $U\s V$ is \emph{$k$-balanced} if there exists a min-cut whose two sides $S_1,S_2$ satisfy $|S_i\cap U|\ge k$ for both $i=1,2$. More specifically, we say that $U$ is \emph{$k$-balanced with witness $(S_1,S_2)$}.
\ED
By definition, a subset $U\s V$ is either $k$-unbalanced or $k$-balanced (or possibly both, if there are multiple min-cuts in the graph). 

We now briefly describe our algorithm. If $U$ is $k$-unbalanced with witness $S$, then the algorithm computes a family $\m F$ of subsets of $U$ of size $k^{O(1)}\polylog(n)=\pl(n)$ such that some subset $R\in\m F$ satisfies $|R\cap S|=1$. The algorithm then executes \thm{small} on each subset in $\m F$, guaranteeing that the target set $R$ is processed and the min-cut is found. 
Otherwise, $U$ must be $k$-balanced with some witness $S$, and the algorithm computes a subset $U'\s U$ such that $|U'|\le|U|/2$ and both $S\cap U'\ne\emptyset$ and $(V\sm S)\cap U'\ne\emptyset$. Of course, the algorithm does not know which case actually occurs, so it executes both branches. But the second branch can only happen $O(\logn)$ times before $|U|\le k$, at which point we can simply run $s$--$t$ min-cut between all pairs of vertices in $U$.

\begin{algorithm}
\caption{DeterministicMincut$(G=(V,E),T)$} \label{alg:main}
\begin{algorithmic}[1]
\State $U\gets T$
\State $k\gets C\log^Cn$ for a sufficiently large constant $C>0$
\While{ $|U|\ge k$ }
 \State Run \thm{small} on $U$ \Comment{Handles case when $U$ is $k$-unbalanced (see \defn{balanced})}\linel{small}
 \State Compute $U'$ from $U$ according to \thm{large} \Comment{Handles case when $U$ is $k$-balanced}
 \State Update $U\gets U'$ \Comment{$|U|$ shrinks by at least factor $2$}
\EndWhile
\For{each pair of distinct $s,t\in U$}
 \State Compute min $s$--$t$ cut in $G$\linel{base}
\EndFor
\State\Return smallest cut seen in lines~\ref{line:small}~and~\ref{line:base}
\end{algorithmic}
\end{algorithm}

\subsection{Unbalanced Case}\secl{hash}

In this section, we solve the case when $U$ is $k$-unbalanced (\line{small}) for some fixed $k=\pl(n)$.

\begin{restatable}[Unbalanced case]{theorem}{Small}\thml{small}
Consider a graph $G=(V,E)$, a parameter $k\ge1$, and a $k$-unbalanced set $U\s T$. Then, we can compute the Steiner min-cut in $k^{O(1)}\pl(n)$ many $s$--$t$ max-flow computations plus $\tO(m)$ deterministic time.
\end{restatable}

\subsubsection{Unbalanced Case:\ De-randomization}

Recall from the \textbf{Unbalanced Case} of \sec{techniques} that our goal is to de-randomize thes random process of sampling each vertex independently with probability $1/k$. We compute a deterministic family of subsets $R\s T$ such that for any subset $S$ of size at most $k$ (in particular, for the set witnessing the fact that $U$ is $k$-unbalanced), there exists a subset $R$ in the family with $|R\cap S|=1$.

\BT\corl{isolator}
For every $n$ and $k<n$, there is a deterministic algorithm that constructs a family $\m F$ of subsets of $[n]$ such that, for each subset $S\s[n]$ of size at most $k$, there exists a set $S'\in\m F$ with $|S\cap S'|=1$. The family $\m F$ has size $k^{O(1)}\logn$ and contains no sets of size at most $1$, and the algorithm takes $k^{O(1)}n\logn$ time.
\ET
Before we prove \cor{isolator}, we first show why it implies an algorithm for the unbalanced case as promised by \thm{small}, restated below.
\Small*
\BP
Let $S$ be the set witnessing the fact that $U$ is $k$-unbalanced.
Apply \cor{isolator} with parameters $n=|U|$ and $k$. Map the elements of $[n]$ onto $U$, obtaining a family $\m F$ of subsets of $U$ such that for any set $S'\s U$ with $|S'|\le k$, there exists a set $R\in\m F$ with $|R|\ge2$ and $|R\cap S'|=1$. In particular, for the set $S'=S\cap U$, we have $1=|R\cap S'|=|R\cap (S\cap U)|=|R\cap S|$. Invoke \thm{local} on the set $R$ to obtain, for each $v\in R$, a set $S_v$ satisfying $S_v\cap R=\{v\}$ that minimizes $w(\pt S_v)$, along with the value $w(\pt S_v)$. Finally, output the set $S_v$ with minimum value of $w(\pt S_v)$. To show that $S_v$ is a min-cut of graph $G$, it suffices to verify that $S_v$ is a valid cut (that is, $\emptyset\subsetneq S_v\subsetneq V$), and that $w(\pt S_v)\le w(\pt S)$.

Since $|R|\ge2$, the set $S_v$ satisfies $\emptyset\subsetneq S_v\subsetneq R$, so it is a cut of the graph $G$. 
Since $|R\cap S|=1$, for the vertex $u\in U$ with $R\cap S=\{u\}$, the set $S$ satisfies the constraints for $S_u$. In particular, $w(\pt S_u)\le w(\pt S)$. We output the set $S_v$ minimizing $w(\pt S_v)$, so $w(\pt S_v)\le w(\pt S_u)\le w(\pt S)$, as promised.
\EP

The rest of this section focuses on proving \cor{isolator}. We first prove an easier variant, where we allow sets of size at most $1$.

\BT\thml{isolator}
For every $n$ and $k$, there is a deterministic algorithm that constructs a family $\m F$ of subsets of $[n]$ such that, for each subset $S\s[n]$ of size at most $k$, there exists a set $S'\in\m F$ with $|S\cap S'|=1$. The family $\m F$ has size $k^{O(1)}\logn$ and the algorithm takes $k^{O(1)}n\logn$ time.
\ET
To prove \thm{isolator}, we use the following de-randomization building block due to \cite{alon1995color}. The theorem below is from \cite{FPT-book}, who state it in terms of \emph{$(n,k,k^2)$-splitters} (which we will not define here for simplicity).
\BT[Theorem 5.16 from \cite{FPT-book}]\thml{splitter}
For any $n,k\ge1$, one can construct a family of functions from $[n]$ to $[k^2]$ such that for every set $S\s[n]$ of size $k$, there exists a function $f$ in the family whose values $f(i)$ are distinct over all $i\in S$. The family has size $k^{O(1)}\logn$ and the algorithm takes time $k^{O(1)}n\logn$.
\ET
\BP[Proof of \thm{isolator}]
Apply \thm{splitter} to $n$ and $k$, and for each function $f:[n]\to[k^2]$ in the constructed family, add the sets $f\inv(j)$ for all $j\in[k^2]$ to our family $\m F$ of subsets of $[n]$. Fix any set $S\s[n]$ of size $k$. For the function $f$ guaranteed by \thm{splitter} for this set $S$, we have $|f\inv(f(i)) \cap S|=1$ for any $i\in S$. Therefore, setting $S'=f(i)$ for any $i\in S$ suffices.

This only handles subsets $S\s[n]$ of size \emph{exactly} $k$, but we can repeat the above construction for each positive integer $k'\le k$. The total size and running time go up by a factor of $k$, which is absorbed by the $k^{O(1)}$ factors.
\EP

Finally, to prove \cor{isolator}, we add the condition that $\m F$ cannot contain sets of size at most $1$, at the cost of imposing the additional constraint $k<n$.

\BP[Proof of \cor{isolator}]
The only difference in the output is that $\m F$ must contain no sets of size at most $1$. Apply \thm{isolator} to $n$ and $k$ to obtain a family $\m F_0$. Initialize a set $\m F$ as $\m F_0$ minus all subsets of size at most 1. For each singleton set $\{x\}\in\m F_0$, choose $k$ arbitrary elements in $[n]\sm x$, and for each chosen element $y$, add the set $\{x,y\}$ to $\m F$. The total size of $\m F$ increases by at most a factor $k$. Now consider a subset $S\s[n]$ of size at most $k$, and let $S'$ be a set in $\m F_0$ with $|S\cap S'|=1$, as promised by \thm{isolator}. If $|S'|>1$, then $S'\in\m F$ as well. Otherwise, if $S'=\{x\}$, then since $|S\sm x|<k$ and we chose $k$ elements $y\in[n]\sm x$, there exists some chosen $y\notin S$ for which $\{x,y\}$ was added to $\m F$. This set $\{x,y\}$ satisfies $|S\cap \{x,y\}|=1$.
\EP

\subsection{Balanced Case: Sparsifying \textit{U}}

If $U$ is $k$-balanced, then, as mentioned in the \textbf{Balanced Case} of \sec{techniques}, we compute a subset $U'\s U$ of size at most $|U|/2$ using \emph{expander decompositions}. This section is dedicated to proving the theorem below.
\BT[Sparsification of $U$]\thml{large}
Fix any constant $\e>0$. Then, there is a constant $C>0$ (depending on $\e$) such that the following holds. Consider a graph $G=(V,E)$, a parameter $\phi\le1/(C\log^Cn)$, and a set $U\s V$ of vertices that is $(1+1/\phi)^3$-balanced with witness $(S_1,S_2)$. Then, we can compute in deterministic $O(m^{1+\e})$ time  a set $U'\s U$ with $|U'|\le|U|/2$ such that $S_i\cap U'\ne\emptyset$ for both $i=1,2$.
\ET

\subsubsection{Expanders and Expander Decomposition}

Given a graph $G=(V,E,w)$, we first introduce some notation. For disjoint vertex subsets $V_1,\lds,V_\el\s V$, define $E(V_1,\lds,V_\el)$ as the set of edges $(u,v)\in E$ with $u\in V_i$ and $v\in V_j$ for some $i\ne j$. Recall that $w(F)$ is the sum of weights of edges in $F$; i.e., $w(E(V_1,\lds,V_\el))$ is the sum of weights of edges with endpoints in different vertex sets in $V_1, V_2, \ldots, V_\el$. In particular, for a cut $(A, B)$, we denote the edges in the cut both by $E(A, B)$ as well as the previously introduced notation $\pt A$ (or $\pt B$), and the weight of the cut is correspondingly denoted $w(E(A, B))$ as well as $w(\pt A)$ (or $w(\pt B)$).
 For a vector $\bd\in\R^V$ of entries on the vertices, define $\bd(v)$ as the entry of $v$ in $\bd$, and for a subset $U\s V$, define $\bd(U):=\sum_{v\in U}\bd(v)$.

We now introduce the concept of an expander ``weighted'' by \emph{demands} on the vertices. 

\BD[$(\phi,\bd)$-expander]
Consider a weighted graph $G=(V,E,w)$ and a vector $\bd\in\R^V_{\ge0}$ of nonnegative entries on the vertices (the ``demands''). The graph $G$ is a \emph{$(\phi,\bd)$-expander} if for all subsets $S\s V$,
\[ \f{w(\pt S)}{\min\{\bd(S),\bd(V\sm S)\}} \ge \phi.\]
\ED

Intuitively, to capture the intersection of a set with $U$, we will place demand $\la$ at each vertex $v\in U$, where $\la$ is the weight of the min-cut, and demand $0$ at the remaining vertices. We now present a deterministic algorithm that computes our desired expander decomposition.


\begin{restatable}[$(\phi,\bd)$-expander decomposition algorithm]{theorem}{Exp}\thml{exp}
Fix any constant $\e>0$ and parameter $0<\phi\le(\log n)^{-O(1/\e^4)}$. Given a weighted graph $G=(V,E,w)$ and a demand vector $\bd\in\R^V_{\ge0}$ of nonnegative, polynomially-bounded entries on the vertices, there is a deterministic algorithm running in $O(m^{1+\e})$ time that partitions $V$ into subsets $V_1,\lds,V_\el$ such that
 \BE
 \im For each $i\in[\el]$, define the demands $\bd_i\in\R^{V_i}_{\ge0}$ as $\bd_i(v)=\bd(v)+w(E(\{v\},V\sm V_i))$ for all $v\in V_i$. Then, the graph $G[V_i]$ is a $(\phi,\bd_i)$-expander.
 \im The total weight $w(E(V_1,\lds,V_\el))$ of inter-cluster edges is $(\log n)^{O(1/\e^4)}\phi\, \bd(V)$.
 \EE
\end{restatable}
The theorem is almost identical to Corollary~2.5 of~\cite{LS21}, except that $\bd_i(v)=\bd(v)+w(E(\{v\},V\sm V_i))$ instead of $\bd_i(v)=\bd(v)$. For completeness, we provide a proof of \thm{exp} in \sec{exp} which uses Corollary~2.5 of~\cite{LS21} as a black box.

\subsubsection{Sparsification Algorithm}

Let $\tilde\la \in [\la,3\la]$ be a 3-approximation to the min-cut $\la$, which can be computed in deterministic $\tO(m)$ time using the $(2+\de)$-approximation algorithm of Matula (for any $\de>0$)~\cite{matula1993linear}. Set $\phi:=1/(C\log^Cn)$ for a sufficiently large constant $C>0$, and let $\e>0$ be the constant fixed by \thm{main}. We apply \thm{exp} to $G$ with parameters $\e,\phi$ and the demand vector $\bd\in\R^V_{\ge0}$ satisfying $\bd(v)=\tilde\la$ for all $v\in U$ and $\bd(v)=0$ for all $v\in V\sm U$. Observe that $\bd(V)=|U|\cd\tilde\la\le|U|\cd3\la$. Let $V_1,\lds,V_\el\s V$ be the output, and for each $i\in[\el]$, define $U_i:=V_i\cap U$.

We now describe the procedure to select the subset $U'\s U$. Call each cluster $V_i$ \emph{trivial} if $U_i=\emptyset$, \emph{small} if $1\le|U_i|\le1/\phi^2$, and \emph{large} if $|U_i|>1/\phi^2$. The algorithm for selecting the set $U'$ is simple: 
\begin{itemize}
\item[--] for each trivial cluster, do nothing;
\item[--] for each small cluster $V_i$, add an arbitrary vertex of $U_i$ to $U'$;
\item[--] for each large cluster $V_j$, add $1+1/\phi$ arbitrary vertices of $U_j$ to $U'$. 
\end{itemize}

\subsubsection{Size Bound}

First, we prove the desired size bound of the sparsified set $U'$, which is one part of \thm{large}.

\BCL\clml{num-clusters}
There are at most $\tO(\phi |U|)$ many clusters; that is, $\el\le\tO(\phi |U|)$.
\ECL
\BP
Since $\la$ is the min-cut of graph $G$, each cluster $V_i$ has $w(\pt V_i)\ge\la$, so the total weight of inter-cluster edges is at least $\el \la/2$. By the guarantee of \thm{exp}, the total weight of inter-cluster edges is at most $\tO(\phi \,\bd(V))=\tO(\phi |U|\tilde \la)\le \tO(\phi |U|\la)$. Putting these together gives $\el\le\tO(\phi|U|)$.
\EP

\BC\corl{size}
There exists a constant $C>0$ (depending on $\e$) such that if $\phi\le1/(C\log^Cn)$, then the set $U'$ constructed by the sparsification algorithm satisfies $|U'|\le|U|/2$.
\EC
\BP
There are at most $\tO(\phi|U|)$ small clusters by \clm{num-clusters}, and there are at most $\phi^2|U|$ large clusters.
This gives 
\[ |U'|\le \tO(\phi |U|)+\phi^2|U|\cd(1+1/\phi) \le \tO(\phi |U|) \le \phi|U| \cd \f C2\log^Cn  \]
for an appropriate constant $C>0$ (depending on $\e$). If $\phi\le1/(C\log^Cn)$, then 
\[ |U'| \le \phi|U| \cd \f C2\log^Cn \le |U|/2 .\]
\EP

\subsubsection{Hitting Both Sides of the Min-cut}

In this section, we prove the ``hitting'' property of the sparsified set $U'$ in \thm{large}, namely the guarantee that $S_i\cap U'\ne\emptyset$ for both $i=1,2$.

The claim below says that the min-cut $(A,B)$ cannot cut too ``deeply'' into the sets $U_i$. In particular, if a set $U_i$ is large (say, $|U_i| \gg 1/\phi$), then the min-cut cannot cut $U_i$ evenly in the sense that $|U_i\cap A|\approx|U_i\cap B|$; instead, we either have $|U_i\cap A|\ll|U_i\cap B|$ or $|U_i\cap A|\gg|U_i\cap B|$.

\BCL\clml{cutAB}
For any cut $(A,B)$ of $G$, we have
\[ \sum_{i\in[\el]}\min\{|U_i\cap A|,|U_i\cap B|\}\le \f{w(E(A,B))}{\phi\la} ,\]
where $U_i := V_i\cap U$ for $i\in [\ell]$.
\ECL
\BP
Since $G[V_i]$ is a $(\phi, \bd_i)$-expander, and since $\bd_i(S)\ge\bd(S)=|U\cap S|\cd\tilde\la\ge|U\cap S|\cd\la$ for all subsets $S\s V_i$, we have
\[ \f{w(E(V_i\cap A,V_i\cap B))}{\min\{|U\cap(V_i\cap A)|\cd\la,|U\cap(V_i \cap B)|\cd\la\} } \ge \f{w(E(V_i\cap A,V_i\cap B))}{\min\{\bd_i(U_i\cap A),\bd_i(U_i\cap B)\} } \ge \phi ,\]
which means that
\[ \min\{|U_i\cap A|\cd\la,|U_i \cap B|\cd\la\}=\min\{|U\cap(V_i\cap A)|\cd\la,|U\cap(V_i \cap B)|\cd\la\}\le\f{w(E(U_i\cap A,U_i\cap B))}\phi .\]
Since $E(V_i\cap A,V_i\cap B)$ is contained in $E(A,B)$ and is disjoint over all $i$, we have $$\sum_{i\in[\el]} w(E(V_i\cap A,V_i\cap B))\le w( E(A,B)).$$ 
Putting things together,
\[ \sum_{i\in[\el]} \min\{|U_i\cap A|,|U_i\cap B|\} \le\f1\la \sum_{i\in[\el]} \f{w(E(V_i\cap A,V_i\cap B))}\phi\le\f{w(E(A,B))}{\phi\la} .\]
\EP

The next claim states that the min-cut can only cut a few clusters $V_i$ in the sense that both sides of the min-cut intersect $V_i$, This implies that for the sets $U_i\s V_i$ in particular, all but a few of them actually satisfy $U_i\cap A=\emptyset$ or $U_i\cap B=\emptyset$.

\BCL\clml{cutV}
Let $C$ be one side of a min-cut (i.e., $w(\pt C)=\la$). Then, $C$ cuts at most $(1+1/\phi)$ clusters. (We say that $C$ \emph{cuts} cluster $V_i$ if both $C\cap V_i$ and $V_i\sm C$ are non-empty.) 
\ECL
\BP
Suppose for contradiction that $C$ cuts more than $(1+1/\phi)$ clusters. Fix a cluster $V_i$ that is cut, and let $A_i$ and $B_i$ be $C\cap V_i$ and $V_i\sm C$ (possibly swapped) so that $w(E(A_i,V\sm V_i)) \le w(E(B_i,V\sm V_i))$. The edges $E(A_i,B_i)$ are contained in $\pt C$, and across different clusters $V_i$ that are cut, the edges $E(A_i,B_i)$ are disjoint, so
\[ \sum_iw(E(A_i,B_i)) \le w(\pt C) = \la .\]
Since $C$ cuts more than $(1+1/\phi)$ clusters, there exists a cluster $V_i$ with 
\[ w(E(A_i,B_i)) < \f{w(\pt C)}{1+1/\phi}=\f\la{1+1/\phi} .\]
For all subsets $S\s V_i$, we have 
\[\bd_i(S)\ge\sum_{v\in S}w(E(\{v\},V\sm V_i)) = w(E(S,V\sm V_i)) .\]
Since $G[V_i]$ is a $(\phi,\bd_i)$-expander,
\BAL
w(E(A_i,B_i))   &\ge \phi\cd\min\{ \bd_i(A_i) , \bd_i(B_i)\}
\\&\ge \phi \cd\min\{w(E(A_i,V\sm V_i)),w(E(B_i.V\sm V_i))\} 
\\&=\phi\cd w(E(A_i,V\sm V_i)). 
\EAL
Consider the cut $\pt A_i$, which satisfies
\[  w(\pt A_i)=w(E(A_i,B_i))+w(E(A_i,V\sm V_i)) \le w(E(A_i,B_i))+\f1\phi w(E(A_i,B_i))=\lp1+\f1\phi\rp w(E(A_i,B_i))<\la ,\]
contradicting the fact that $C$ is the min-cut.
\EP

Finally, we prove the ``hitting'' property of the sparsified set $U'$. This, along with \cor{size}, finishes the proof of \thm{large}.

\BL
Suppose that $U$ is $(1+1/\phi)^3$-balanced with witness $(S_1,S_2)$. Then, for the set $U'$ constructed by the sparsification algorithm, we have $S_i\cap U'\ne\emptyset$ for both $i=1,2$.
\EL

\BP
For each cluster $V_i$, by \clm{cutAB},
\[ \min\{|U_i\cap A|,|U_i\cap B|\}\le \f{ w(E(A,B))}{\phi\la} \le\f{1}\phi .\]
In other words, either $|S_1\cap U_i|\le 1/\phi$ or $|S_2\cap U_i|\le1/\phi$. Call a cluster $V_i$:
 \BE
 \im \emph{white} if $S_1\cap U_i=\emptyset$ (i.e., $U_i\s S_2$).
 \im \emph{light gray} if $0<|S_1\cap U_i|\le |S_2\cap U_i|<|U_i|$, which implies that $0<|S_1\cap U_i|\le1/\phi$.
 \im \emph{dark gray} if $0<|S_2\cap U_i|<|S_1\cap U_i|<|U_i|$, which implies that $0<|S_2\cap U_i|\le1/\phi$.
 \im \emph{black} if $S_2\cap U_i=\emptyset$ (i.e., $U_i\s S_1$).
 \EE
Every cluster must be one of the four colors, and by \clm{cutV}, there are at most $(1+1/\phi)$ many (light or dark) gray clusters since $U_i\cap S_1, U_i\cap S_2 \not= \emptyset$ implies that $S_1$ cuts cluster $V_i$. Note that since we are only considering clusters $V_i$ such that $U_i \not= \emptyset$, it must be that for a white cluster, we have $|S_2\cap U_i| \not= \emptyset$, and similarly, for a black cluster, we have $|S_1\cap U_i| \not= \emptyset$. There are now a few cases:
 \BE
 \im There are no large clusters. In this case, if there is at least one white and one black small cluster, then the vertices from these clusters added to $U'$ are in $S_2$ and $S_1$, respectively.
 Otherwise, assume w.l.o.g.\ that there are no black clusters. Since there are at most $(1+1/\phi)$ gray clusters in total, $|S_1\cap U|\le (1+1/\phi)\cd1/\phi^2$, contradicting our assumption that $\min\{|S_1\cap U|,|S_2\cap U|\}\ge(1+1/\phi)^3$. 
 \im There are large clusters, but all of them are white or light gray. Let $V_i$ be a large white or light gray cluster. Since we select $1+1/\phi$ vertices of $U_i$, and $|S_1\cap U_i|=\min\{|S_1\cap U_i|,|S_2\cap U_i|\}\le 1/\phi$, we must select at least one vertex not in $S_1$. Therefore, $S_2\cap U'\ne\emptyset$. If there is at least one black cluster, then the selected vertex in there is in $U'$, so $S_1\cap U'\ne\emptyset$ too, and we are done.
 
 So, assume that there is no black cluster. Since all large clusters are light gray (or white), $|S_1\cap U_i| \le 1/\phi$ for all large clusters $V_i$. Moreover, by definition of small clusters, $|S_1\cap U_i| \leq |U_i| \le 1/\phi^2$ for all small clusters $V_i$. Since there are at most $(1+1/\phi)$ gray clusters by \clm{cutV},
\BAL
|S_1\cap U| &= \sum_{i: V_i\text{ small}}|S_1\cap U_i| + \sum_{i: V_i\text{ large}}|S_1\cap U_i| \\&\le \lp1+\f1\phi\rp\cd\f1{\phi^2} + \lp1+\f1\phi\rp\cd\f1\phi = 2\lp1+\f1\phi\rp\cd\f1\phi < \lp1+\f1\phi\rp^3 ,
\EAL
a contradiction.
\im There are large clusters, but all of them are black or dark gray. Symmetric case to (2) with $S_1$ replaced with $S_2$.
\im There is at least one black or dark gray large cluster $V_i$, and at least one white or light gray large cluster $V_j$. In this case, since we select $1+1/\phi$ vertices of $U_i$ and $|S_2\cap U_i|=\min\{|S_1\cap U_i|,|S_2\cap U_i||\}\le 1/\phi$, we must select at least one vertex in $S_1$. Similarly, we must select at least one vertex in $U_j$ that is in $S_2$.
 \EE
\EP

\section{Conclusion}
We gave a deterministic algorithm for finding a minimum cut in undirected graphs that uses $O(\log^{O(1)} n)$ calls to any maximum flow algorithm. Using the current best deterministic maximum flow algorithms, this yields an overall running time of $\tilde O(m \cdot \min(\sqrt{m}, n^{2/3}))$ for weighted graphs, 
    and $m^{4/3+o(1)}$ for unweighted (multi)-graphs. This marks the first improvement for this problem since a running time bound of $\tO(mn)$ was established by several papers in the early 1990s. 
    
    Our result is obtained as an application of a new technique that we call isolating cuts. Our main observation is that, given a subset of vertices called terminals, using $O(\log n)$ maximum flow calls, we can find the minimum cuts separating each individual terminal from the rest of the terminals. This immediately yields a simple randomized minimum cut algorithm, and our eventual deterministic algorithm can be viewed as a derandomization of this randomized algorithm. In fact, we obtain the same running time for the more general Steiner connectivity problem, where we are given a subset of terminals and need to find the minimum weight cut with at least one terminal on each side of the cut. For this latter problem, our algorithm is an improvement on even the best randomized algorithm that was previously known.
    
    The immediate open problem suggested by our result is an $m^{1+o(1)}$-time deterministic minimum cut algorithm, which has already been obtained by Li~\cite{Li21} since the first publication of our work. We believe the isolating cuts technique can be a crucial component in solving other longstanding questions in graphs algorithms as well. One particularly fascinating question is to break the existing 60-year old barrier for the all-pairs minimum cuts problem. In spite of much effort, the state of the art for this latter problem (on general, weighted graphs) remains the classic 1961 algorithm of Gomory and Hu that reduces it to $n-1$ maximum flow calls. It is entirely plausible, however, that this problem can actually be solved using just $O(\log^{O(1)} n)$ maximum flow calls, and we believe the isolating cuts technique can be a valuable technical tool for this purpose.
    
\section*{Acknowledgements}    

JL was supported in part by NSF award CCF-1907820. DP was supported in part by NSF award CCF-1955703 and an NSF CAREER award CCF-1750140. DP would like to thank David Karger who first introduced him to the deterministic minimum cut problem a decade ago.

  \bibliographystyle{alpha}
\bibliography{refs,dp-refs}

\appendix

\section{Weighted Expander Decomposition}
\label{sec:exp}
In this section, we prove \thm{exp}, restated below.
\Exp*

As discussed right below the statement of \thm{exp}, we use Corollary~2.5 of~\cite{LS21} as a black box. It is identical to \thm{exp} except that $\bd_i(v)=\bd(v)$ instead of $\bd(v)+w(E(\{v\},V\sm V_i))$. To avoid confusion, we use $\bd|_{V_i}$ to denote this new definition.

\BT[Corollary~2.5 of~\cite{LS21}] \thml{LS}
Fix any constant $\e>0$ and any parameter $\phi>0$. Given a weighted graph $G=(V,E,w)$ and a demand vector $\bd\in\R^V_{\ge0}$ of nonnegative, polynomially-bounded entries on the vertices, there is a deterministic algorithm running in $O(m^{1+\e})$ time that partitions $V$ into subsets $V_1,\lds,V_\el$ such that
 \BE
 \im For each $i\in[\el]$, define the demands $\bd|_{V_i}\in\R^{V_i}_{\ge0}$ as $\bd$ restricted to $V_i$: $\bd|_{V_i}(v)=\bd(v)$ for all $v\in V_i$. Then, the graph $G[V_i]$ is a $(\phi,\bd|_{V_i})$-expander.
 \im The total weight $w(E(V_1,\lds,V_\el))$ of inter-cluster edges is $(\log n)^{O(1/\e^4)}\phi\, \bd(V)$.
 \EE
\ET

We also need the flow subroutine below as a ``trimming'' step, following the expander decomposition framework of \cite{SaranurakW19}. It is identical to Theorem~1.5 of~\cite{LiNPS21} with the setting $\e=1/2$, except that paper phrases the result in terms of \emph{fair cuts}; for simplicity, we do not define the concept here.\footnote{In our application, it is enough to compute the expander decomposition in max-flow time, so it suffices to prove \lem{flow} in max-flow time as well, which is trivial. However, we might as well black-box the theorem from~\cite{LiNPS21} for the improved running time.}
\BL\leml{flow}
Given a weighted graph $G=(V,E)$ and two distinct vertices $s,t\in V$, we can find in deterministic $m^{1+o(1)}$ time a $2$-approximate $s$--$t$ mincut $S\s V$ ($s\in S$, $t\notin S$) and a feasible $s$--$t$ flow $f$ such that for each edge $e\in\pt S$, flow $f$ sends at least $1/2$ fraction of the capacity of $e$ in the direction from $S$ to $V\sm S$.
\EL

We now prove \thm{exp}. Apply \thm{LS} to obtain a partition $V_1,\lds,V_\el$ with\linebreak $w(E(V_1,\lds,V_\el)) \le \al\phi\bd(V)$ for some $\al=(\log n)^{O(1/\e^4)}$. For each $V_i$ with $\bd(V_i)\le 2\bd(V)/3$, we recursively apply the algorithm on graph $G[V_i]$ with demands $\bd(v)+w(E(\{v\},V\sm V_i))$. Note that $\bd_i(V_i)=\bd(V_i)+w(\pt V_i) \le 2\bd(V)/3+\al\phi\bd(V)\le3\bd(V)/4$ for small enough $\phi\ll1/\al$, so we make progress with respect to total demand.

If there is a (unique) $V_i$ with $\bd(V_i)>2\bd(V)/3$, then we ``trim'' it as follows. Add a source vertex $s$, and for each vertex $v\in V_i$ with $E(v,V\sm V_i)\ne\emptyset$, add an edge $(s,v)$ of weight $\f1{12\al} w(E(v,V\sm V_i))$. Add a new vertex $t$, and for each vertex $v\in V_i$ with $\bd(v)>0$, add an edge $(v,t)$ of weight $\f\phi2\bd(v)$. Call \lem{flow} on graph $G_i$, and let $S_i\s V_i\cup\{s\}$ be the output. The key claim is that for the new ``trimmed'' cluster $V_i'=V_i\sm S_i$, the graph $G[V_i']$ is a $(\phi,\bd_i')$-expander for $\bd_i'(v)=\bd(v)+w(E(\{v\},V\sm V_i'))$. In other words, we do not need to recursive on $V'_i$.

\BCL\clml{trim}
We have $\bd(V_i\sm V'_i)\le\bd(V)/4$ and $G[V'_i]$ is a $(\phi/6,\bd_i')$-expander for $\bd_i'(v)=\bd(v)+w(E(\{v\},V\sm V_i'))$.  
\ECL

Therefore, we only need to recursively call the algorithm on the connected components of $G[V_i\sm V'_i]$. Namely, for each connected component $V'$ of $G[S_i\sm\{s\}]$, we call the algorithm on $G[V']$ with demands $\bd(v)+w(E(\{v\},V\sm V'))$. Note that the total demand in this recursive call is $\bd(V')+w(\pt_GV') \le \bd(V_i\sm V'_i) + w(\pt V_i) + w(\pt_{G_i}S_i)$. We have $\bd(V_i\sm V'_i)\le\bd(V)/4$, and by \lem{flow}, the cut $S_i$ is a $3$-approximate $s$--$t$ mincut in $G_i$, so its weight $w(\pt_{G_i}S_i)$ has weight at most $\f\phi4\bd(V)$ since $\{s\}$ is a valid $s$--$t$ cut of weight $\f1{12\al}w(\pt_GV_i)\le\f1{12\al} \cd \al\phi\bd(V)=\f\phi{12}\bd(V)$. The total demand is therefore at most $\bd(V)/4 + \al\phi\bd(V) + \f\phi4\bd(V)$, which is at most $\bd(V)/2$ for $\phi$ small enough.

Since all demands and weights are polynomially bounded, and since each recursive call has total demand a constant fraction smaller, the recursion depth is $O(\log n)$. The final expander decomposition satisfies the given requirements except that $\phi$ is replaced by $\phi/6$, but we can always re-parameterize $\phi$ accordingly and only lose constant factors everywhere.

It remains to prove \clm{trim}. For the first statement $\bd(V_i\sm V'_i)\le\bd(V)/4$, observe that for each vertex $v\in V_i\sm V'_i$ with $\bd(v)>0$, the edge $(v,t)$ of weight $\f\phi2\bd(v)$ is cut. Therefore, $\f\phi2\bd(V_i\sm V'_i)\le w(\pt_{G_i}S_i)$, which we already argued is at most $\f\phi4\bd(V)$ for small enough $\phi$, as desired. For the second statement, suppose for contradiction that $G[V'_i]$ is not a $(\phi/6,\bd'_i)$-expander. Then, there is a cut $U\s V'_i$ with $w(\pt_{G[V'_i]}U)\le \f\phi6\bd'_i(U) = \f\phi6(\bd(U) + w(E(U,V\sm V'_i)))$. Since $G[V_i]$ is a $(\phi,\bd|_{V_i})$-expander, $w(\pt_{G[V_i]}U)\ge\phi\bd|_{V_i}(U)=\phi\bd(U)$. Taking the difference of the two inequalities gives $w(E(U,V_i\sm V'_i))=w(\pt_{G[V'_i]}U)-w(\pt_GU)\ge\f{5\phi}6\bd(U)-\f\phi6 w(E(U,V\sm V'_i))$. 

By the properties of \lem{flow} on the flow problem on $G_i$, there is a feasible $s$--$t$ flow $f$ that sends at least $1/2$ fraction of the capacity of each edge $e\in\pt S$ in the direction from $S$ to $V\sm S$. The edges $e\in\pt S$ can be partitioned into three types: the edges $e$ adjacent to $s$, the edges in $G[V_i]$, and the edges adjacent to $t$. Consider the edges of the first two types with (exactly) one endpoint in $U$. These edges have total capacity 
\BALN
 &\f1{12\al} w(E(U,V\sm V_i))+w(E(U,V_i\sm V'_i)) \nonumber
\\ ={} & \f1{12\al} w(E(U,V\sm V_i))+\f15w(E(U,V_i\sm V'_i))+\f45w(E(U,V_i\sm V'_i)) \nonumber
\\ \ge{} & \f1{12\al} w(E(U,V\sm V_i))+\f15w(E(U,V_i\sm V'_i))+\f45\lp\f{5\phi}6\bd(U)-\f\phi6 w(E(U,V\sm V'_i))\rp \nonumber
\\ \ge{} & \f1{12\al} w(E(U,V\sm V_i))+\f15w(E(U,V_i\sm V'_i))+\f{2\phi}3\bd(U)-\f{2\phi}{15} w(E(U,V\sm V'_i)) . \eqnl{source}
\EALN
Flow $f$ sends at least $1/2$ fraction of this capacity from $S$ to $U$, and this flow must eventually reach $t$. It can escape $U$ in two ways: through edges in $G[V'_i]$ and through edges adjacent to $t$. The total capacity of these edges is
\BALN
w(\pt_{G[V'_i]}U)+\f\phi2\bd(U) &\le \f\phi6\lp \bd(U) + w(E(U,V\sm V'_i)) \rp + \f\phi2\bd(U) \nonumber
\\&= \f{2\phi}3\bd(U) + \f\phi6w(E(U,V\sm V'_i)). \eqnl{cap}
\EALN
Using that $w(E(U,V\sm V'_i))=w(E(U,V\sm V_i))+w(E(U,V_i\sm V'_i))$ and comparing term by term, we conclude that for $\phi$ much smaller than $\al$, expression~\eqn{source} multiplied by $1/2$ is strictly larger than expression~\eqn{cap}, which means the flow entering $U$ cannot completely escape $U$, a contradiction.

\end{document}